\documentclass[conference]{IEEEtran}
\IEEEoverridecommandlockouts
\usepackage{amsmath,amssymb,amsfonts}
\usepackage{algpseudocode}
\usepackage[linesnumbered,ruled,vlined]{algorithm2e}
\usepackage{pifont}
\usepackage{graphicx}
\usepackage{textcomp}
\usepackage{xcolor}
\usepackage{pgf}
\usepackage{tikz}
\usepackage{color}
\usepackage{stfloats}
\usepackage{multirow}
\usepackage{diagbox}
\usepackage[hyphens]{url}
\usepackage{subcaption}
\usepackage{adjustbox}
\usepackage{parallel}
\usepackage{float}
\usepackage{placeins}
\usetikzlibrary{shapes.multipart, arrows.meta, positioning, calc}

\usepackage{fancyhdr}
\usepackage[normalem]{ulem}
\usepackage[sort,nocompress]{cite}

\usepackage{booktabs}

\usepackage{comment}
\usepackage[bookmarks=true,breaklinks=true,colorlinks,linkcolor=black,citecolor=blue,urlcolor=black]{hyperref}

\definecolor{darkred}{rgb}{192, 0, 0} 

\newcommand{\blackcircled}[1]{%
  \adjustbox{valign=c}{%
    \tikz[baseline=(char.base)]{
      \node[shape=circle, draw=black, fill=black, inner sep=0.5pt] (char) {\color{white}\sffamily\bfseries #1};
    }%
  }%
}

\newcommand{\squishlist}{
   \begin{list}{$\bullet$}
    { \setlength{\itemsep}{0pt}      \setlength{\parsep}{0pt}
      \setlength{\topsep}{3pt}       \setlength{\partopsep}{0pt}
      \setlength{\listparindent}{-2pt}
      \setlength{\itemindent}{-5pt}
      \setlength{\leftmargin}{1em} \setlength{\labelwidth}{0em}
      \setlength{\labelsep}{0.5em} } }

\newcommand{\squishend}{
    \end{list}  }

\newcommand*\blackcircledempty[1]{\tikz[baseline=(char.base)]{
        \node[shape=circle, text={rgb,255:red,0;green,0;blue,0}, font=\small, draw={rgb,255:red,0;green,0;blue,0},inner sep=0.5pt] (char) {#1};}}

\def\BibTeX{{\rm B\kern-.05em{\sc i\kern-.025em b}\kern-.08em
    T\kern-.1667em\lower.7ex\hbox{E}\kern-.125emX}}
    
\begin{document}

\title{TLV-HGNN: Thinking Like a Vertex for Memory-efficient HGNN Inference \\
\thanks{
This work was supported by the National Key Research and Development Program (No. 2022YFB4501400), the National Natural Science Foundation of China (No. 62202451), CAS Project for Young Scientists in Basic Research (No. YSBR-029), Beijing Nova Program (No. 20250484774), the Postdoctoral Fellowship Program of CPSF (No. GZB20250397) and CAS Project for Youth Innovation Promotion Association.

Mingyu Yan is the corresponding author (e-mail: yanmingyu@ict.ac.cn).
}
}

\vspace{-20pt}
\author{
Dengke Han\textsuperscript{1,2}, 
Duo Wang\textsuperscript{1,2}, 
Mingyu Yan\textsuperscript{1,2}, 
Xiaochun Ye\textsuperscript{1,2}, 
Dongrui Fan\textsuperscript{1,2}\\
\textsuperscript{1 }State Key Lab of Processors, Institute of Computing Technology, CAS\\
\textsuperscript{2 }University of Chinese Academy of Sciences
}



\maketitle
\vspace{-20pt}

\vspace{-60pt}
\begin{abstract}

Heterogeneous graph neural networks (HGNNs) excel at processing heterogeneous graph data and are widely applied in critical domains. In HGNN inference, the neighbor aggregation stage is the primary performance determinant, yet it suffers from two major sources of memory inefficiency. First, the commonly adopted per-semantic execution paradigm stores intermediate aggregation results for each semantic prior to semantic fusion, causing substantial memory expansion. Second, the aggregation process incurs extensive redundant memory accesses, including repeated loading of target vertex features across semantics and repeated accesses to shared neighbors due to cross-semantic neighborhood overlap. These inefficiencies severely limit scalability and reduce HGNN inference performance.

In this work, we first propose a semantics-complete execution paradigm from a vertex perspective that eliminates per-semantic intermediate storage and redundant target vertex accesses. Building on this paradigm, we design TVL-HGNN, a reconfigurable hardware accelerator optimized for efficient aggregation. In addition, we introduce a vertex grouping technique based on cross-semantic neighborhood overlap, with hardware implementation, to reduce redundant accesses to shared neighbors.  Experimental results demonstrate that TVL-HGNN achieves average speedups of 7.85$\times$ and 1.41$\times$ over the NVIDIA A100 GPU and the state-of-the-art HGNN accelerator HiHGNN, respectively, while reducing energy consumption by 98.79\% and 32.61\%.


\end{abstract}

\section{Introduction}


Graph Neural Networks (GNNs) have shown remarkable capability in modeling non-Euclidean data, driving their adoption in diverse domains. Early research predominantly focused on homogeneous graphs (HomoGs), consisting of a single vertex and edge type. However, real-world data is often heterogeneous, giving rise to heterogeneous graphs (HetGs) that contain multiple vertex and edge types. Unlike GNNs for HomoGs, Heterogeneous Graph Neural Networks (HGNNs) are designed to capture both structural dependencies and the rich semantic information from diverse relation types. This enhanced representational capacity has made HGNNs essential in various critical domains, including recommendation systems~\cite{weibo-recommendation, dataset-recommendation}, medical analysis~\cite{medical_analysis, medical_analysis_aaai}, and beyond~\cite{Comprehensive_Survey_GNN_Distributed_Training}.


The inference process of mainstream HGNN models~\cite{R-GCN, R-GAT, NARS} generally comprises four stages: \textit{Semantic Graph Build} (SGB), \textit{Feature Projection} (FP), \textit{Neighbor Aggregation} (NA), and \textit{Semantic Fusion} (SF), each with distinct execution behaviors. SGB extracts neighbors according to semantic types; FP performs matrix multiplications; NA traverses graph-structured neighbors to execute element-wise aggregations; and SF conducts aggregation across semantic-specific results. Among these, NA dominates execution time and constitutes the primary performance bottleneck during inference~\cite{understand_HGNN, understand_hgnn_training}.



Despite its central role in HGNN inference, the NA stage is also the primary source of memory inefficiency under the conventional paradigm. \textbf{First}, the prevalent per-semantic execution performs neighbor aggregation for each semantic graph independently, followed by semantic fusion to integrate intermediate results and capture semantic information. This necessitates storing outputs for all semantics, causing substantial memory expansion that, as the graph size increases, can often trigger out-of-memory (OOM) failures and severely hinder scalability. \textbf{Second}, NA incurs extensive redundant accesses to both target and source vertex features: target features are repeatedly loaded across semantics, and neighborhood overlaps among targets lead to additional redundant accesses to shared neighbors, collectively degrading inference performance.


To address these limitations, we first propose a semantics-complete execution paradigm from a vertex perspective that mitigates memory expansion and eliminates redundant accesses to target vertices. Then we design a dynamically reconfigurable hardware accelerator, TVL-HGNN, to efficiently support this paradigm. In addition, we introduce an overlap-driven vertex grouping technique to maximize the reuse of shared neighbors across target vertices, thereby reducing DRAM accesses. The main contributions of this work are as follows:\par
\squishlist
\item
We quantitatively analyze HGNN inference and reveal substantial memory expansion and redundant memory accesses.
\item
We propose a semantics-complete inference paradigm thinking like a vertex, and design TVL-HGNN, a reconfigurable accelerator tailored to efficiently support it.
\item 
We present an overlap-driven vertex grouping technique to improve data locality and reduce DRAM accesses.
\item
Extensive experiments demonstrate that TVL-HGNN outperforms the GPU A100 and the state-of-the-art (SOTA) HGNN accelerator HiHGNN~\cite{HiHGNN} in terms of performance and energy efficiency, with superior scalability.

\squishend

\section{Background}

\subsection{Heterogeneous Graph}

A graph is formally defined as $G=(V,E,\mathcal{S}^v,\mathcal{S}^e)$~\cite{Simple-HGN, SeHGNN}, with notations specified in Table~\ref{tb:background_notation}, where $V$ is the set of vertices with a vertex type mapping function $\phi:V\rightarrow\mathcal{S}^v$, and $E$ is the set of edges with an edge type mapping function $\psi:E\rightarrow\mathcal{S}^e$.
Each vertex $v_i{\in}V$ is attached with a vertex type $T_v{=}\phi(v_i){\in}\mathcal{S}^v$. Each edge $e_{u, v}{\in}E\,$ is attached with a relation $R_{u,v}{=}\psi(e_{u,v}){\in}\mathcal{S}^e$, starting from the source vertex $u$ to the target vertex $v$.  A graph is considered heterogeneous when $|\mathcal{S}^v|+|\mathcal{S}^e|>2$, otherwise it is homogeneous.

\begin{table}[!htbp]
\centering
\caption{Notations and corresponding explanations.}
\label{tb:background_notation}
\resizebox{0.46\textwidth}{!}{
\tabcolsep=4pt
\begin{tabular}{cc|cc}
\toprule
Notation                & Explanation                           & Notation                  & Explanation       \\ \midrule
$G$                     & heterogeneous graph                   & $V$                       & vertex set \\
$E$                     & edge set                              & $\mathcal{S}^v$           & vertex type set \\
$\mathcal{S}^e$         & edge type set                         & $u,\,v$                   & vertex \\
e ($e_{u,v}$)           & edge (from $u$ to $v$)                & $R$,\ $r$ & relation (semantic) \\
$T_v$                   & vertex type                         & $N_v$           & neighboring set \\
$h_v$, $z_v$                  & embedding                           & $\sigma$           & activation \\
$W$                   & weight matrix                        & $\alpha$           & edge weight \\

\bottomrule
\vspace{-10pt}
\end{tabular}}
\end{table}

\begin{figure*}[!ht] 
	\centering
	\vspace{-5pt}
	\includegraphics[width=0.94\textwidth]{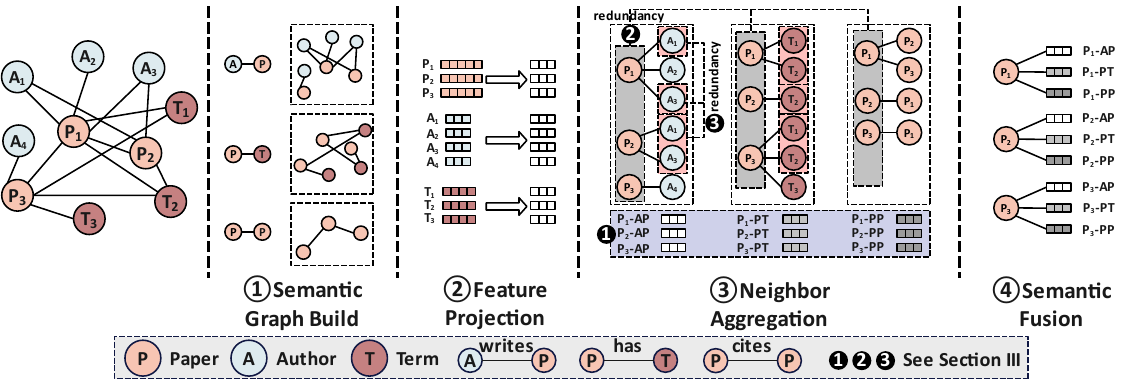}
	\caption{The illustration of HetG and HGNN execution process.}
        \vspace{-14pt}
	\label{fig:background_HetG}
\end{figure*}

Fig.~\ref{fig:background_HetG} presents an example of a HetG from the DBLP dataset, comprising three vertex types: A (Author), P (Paper), and T (Term), along with three relation types: author$\xrightarrow{\rm writes}$paper, paper$\xrightarrow{\rm cites}$paper, and paper$\xrightarrow{\rm has}$term (abbreviated as AP, PP, and PT). Each relation type conveys distinct semantic information between connected vertices and can thus be referred to as a semantic.

\subsection{Heterogeneous Graph Neural Network}

To capture both the structural and semantic information in HetGs, most prevalent HGNN models contain four primary execution stages~\cite{MetaNMP,understand_HGNN,characterization_survey}. \blackcircledempty{1} \textit{Semantic Graph Build (SGB)} stage builds semantic graphs for the following stages by partitioning the original HetG into a set of semantic graphs. 
\blackcircledempty{2} \textit{Feature Projection (FP)} stage transforms the feature vector of each vertex to a new one using a multi-layer perceptron within each semantic graph.
\blackcircledempty{3} \textit{Neighbor Aggregation (NA)} stage performs the aggregation of features from neighbors within each semantic graph.
\blackcircledempty{4} \textit{Semantic Fusion (SF)} stage consolidates the outputs of the NA stage from multiple semantic graphs to obtain unified semantic representations. The general inference process of HGNN models~\cite{R-GCN,R-GAT,NARS} can be formulated as follows using notations in Table~\ref{tb:background_notation}:

$$
\mathbf{z}_v = \sigma \left( \sum_{r \in \mathcal{R}} \sum_{u \in \mathcal{N}_v^r} \alpha_{r, u, v} \mathbf{W}_r \mathbf{h}_u \right)
$$

\subsection{Per-semantic Execution Paradigm}
\label{sec:per-semanticEXE}

Currently, mainstream graph learning frameworks including PyG~\cite{PyG} and DGL~\cite{DGL} all adopt a per-semantic execution paradigm in their HGNN implementations, which has also become the de facto standard for existing HGNN accelerators~\cite{MetaNMP, HiHGNN, ADE-HGNN}. As illustrated in Fig.~\ref{fig:background_HetG}, this paradigm performs neighbor aggregation independently under each semantic, sequentially aggregating the neighbors of all target vertices to capture the structural information of each semantic graph. The resulting intermediate embeddings from all semantics are subsequently fused to generate the final embeddings of target vertices. This paradigm is straightforward to implement and particularly well-suited for leveraging parallel matrix computation, especially on general-purpose GPUs.

\section{Motivation}
\label{sec:motivation}


In this section, we quantitatively identify two memory inefficiencies during HGNN inference: memory expansion and redundant memory access, using the same experimental setup as in Section~\ref{sec:evaluation}. Then we highlight the necessity of thinking like a vertex for achieving efficient HGNN inference.

\subsection{Characterization of HGNN Inference}

Prior work~\cite{understand_HGNN} demonstrates that among all stages of HGNN inference, the NA stage is the most critical, accounting for over 70\% of the total runtime. Its core operation involves traversing the graph to aggregate features from the neighbors of target vertices and exhibits a pronounced memory-bound behavior. Specifically, work~\cite{understand_hgnn_training} reports that the NA stage dominates memory allocation during inference and suffers from a low cache hit rate, leading to both substantial memory consumption and a high volume of DRAM accesses.

\subsection{Significant Memory Expansion}
\begin{figure}[!htbp] 
	\centering
	\vspace{-5pt}
	\includegraphics[width=0.46\textwidth]{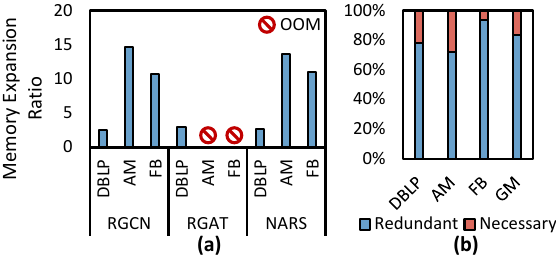}
	\caption{Memory inefficiencies of HGNN inference: (a) Memory expansion; (b) Redundant memory accesses.}
        \vspace{-10pt}
	\label{fig:motivation_memory_inefficiences}
\end{figure}

As detailed in Section~\ref{sec:per-semanticEXE}, the per-semantic execution paradigm necessitates storing intermediate results generated during the NA stage to support subsequent semantic fusion, as illustrated by \blackcircled{1} in Fig.~\ref {fig:background_HetG}, leading to significant memory expansion. Fig.~\ref {fig:motivation_memory_inefficiences}(a) quantifies the \textit{memory expansion ratio defined as the ratio of peak memory usage to the initial memory footprint of the dataset} during inference. As shown, this ratio can reach as high as 15.04 across different models and datasets, occasionally leading OOM errors even on the A100 GPU with 80~GB HBM. Such memory overhead severely limits the scalability of HGNN models when applied to increasingly large real-world graphs. Although this issue can be mitigated through batch-wise execution, doing so significantly degrades inference efficiency~\cite{understand_hgnn_training}.

\subsection{Extensive Redundant Memory Accesses}
\label{sec:motivation_redundant_feature_access}

The NA stage incurs extensive redundant memory accesses during execution, which arise not only from the per-semantic execution paradigm but also from the inherent structural redundancy of HetGs. First, a single target vertex may be involved in multiple semantic graphs, leading to repeated accesses to its feature vector during aggregation, as illustrated by \blackcircled{2} in Fig.~\ref{fig:background_HetG}. Second, target vertices frequently share common neighbors, resulting in duplicated accesses to shared neighbor features, as highlighted by \blackcircled{3} in Fig.~\ref{fig:background_HetG}. Compared to traditional GNNs, these redundancies are significantly amplified in HGNNs due to the multiplicity of semantic views, aggravating memory inefficiencies. As shown in Fig.~\ref{fig:motivation_memory_inefficiences}(b), redundant neighbor feature accesses account for over 80\% of the total feature accesses in geometric mean (GM) across datasets, posing a major bottleneck to inference performance.

\subsection{The Necessity of Thinking Like a Vertex}

\textit{``Thinking like a vertex'' refers to adopting the target vertex and all its neighbors across different semantics as the fundamental aggregation unit, without partitioning the semantic graphs}. Based on the above analysis, two key motivations support a target-vertex-centric approach to HGNN inference. \textbf{First}, memory expansion primarily stems from the fact that each adjacency relation captures only partial semantic information, requiring intermediate aggregation results to be stored until semantic fusion. In contrast, aggregating multi-semantic neighbors directly from the target vertex perspective eliminates the need for such intermediate storage. \textbf{Second}, the structural heterogeneity of semantic graphs implies that even the same pair of target vertices may exhibit inconsistent neighborhood overlap across different semantics, making it difficult to define a unified locality pattern. By shifting to a target-vertex-centric view and jointly considering cross-semantic neighborhood overlap, this challenge can be mitigated while avoiding the overhead of dynamic locality management.


\section{Design}

This section first presents a novel semantics-complete inference paradigm from the perspective of target vertices, designed to mitigate memory expansion and eliminate redundant accesses to target vertex features. Building on this paradigm, we develop TVL-HGNN, a reconfigurable hardware accelerator optimized for efficient execution. Finally, we propose an overlap-driven vertex grouping technique to enhance the reuse of shared neighbor features, thereby reducing DRAM accesses.

\subsection{Semantics-complete Execution}
\label{sec:vertex_centric_execution}

The conventional per-semantic execution paradigm incurs substantial memory expansion and redundant accesses to target vertex features. To address these issues, we propose a novel semantics-complete execution paradigm that adopts a vertex-centric perspective, as detailed in Algorithm~\ref{alg:vertex_centric_paradigm}. In this approach, the neighbors of each target vertex across all semantics are treated as a unified aggregation workload, in contrast to the traditional method that first performs semantic-specific aggregation for all target vertices followed by a separate semantic fusion phase. Specifically, for each target vertex, we first aggregate its neighbors from all semantics (lines 2–8), generating intermediate embeddings for each semantic. These embeddings are then immediately fused to produce the final vertex representation (line 9), leveraging the fact that all required semantic information for that target vertex is already available, thereby eliminating the need for deferred fusion and avoiding the associated data storage and retrieval.






\begin{algorithm}[!tbp]
\SetAlgoLined
\SetKwBlock{DoParallel}{in parallel do}{end}
\caption{Semantics-complete Execution Paradigm}
\label{alg:vertex_centric_paradigm}
\SetKwInOut{Input}{Input}
\SetKwInOut{Output}{Output}
\SetKwInOut{Initial}{Initial}

\Input{Heterogeneous graph $G=(V,E)$, \\
semantics $R = (r_1, r_2, ..., r_n)$, \\
raw features $X = (x_{v1},x_{v2}...x_{u1}..x_{n} )$}
\Output{Vertex embeddings $Z_v=\{z_v|v \in V\}$}
\Initial{$Z_v\ \gets \phi$, $h_{v(u)}' \leftarrow Projection(x_{v(u)}) $}

\For{each target vertices $v \in V$}{
    \For{each semantic $r \in R$}{
        $h_v^r\ \leftarrow h_v'$  \\
        \For{each neighbor $u \in N_v^{r}$}{
            $\alpha_{r,u,v} \leftarrow \text{ComputeEdgeWeight}(h_u',h_v')$ \\
            $h_v^r \leftarrow \text{NeighborAggregate}(h_u', \alpha_{r,u,v})$ \\
        }
    }
    $z_v \leftarrow \text{SemanticFuse}(\{h_v^r|r \in R\})$
}

\Return $Z_v$

\end{algorithm}

This semantics-complete paradigm fundamentally alleviates memory expansion, as it eliminates the need to retain intermediate aggregation results for all target vertices across semantics. Instead, only the intermediate results for a single target vertex need to be stored temporarily and can be discarded once its semantic fusion is performed. Furthermore, by accessing each target vertex only once, the paradigm avoids redundant feature accesses across multiple semantics, considerably reducing DRAM accesses.

\subsection{Hardware Architecture}

\begin{figure}[!ht] 
	\centering
	\vspace{-5pt}
	\includegraphics[width=0.46\textwidth]{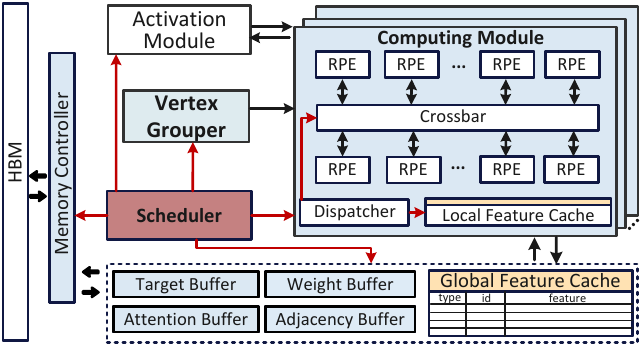}
	\caption{The architecture of TVL-HGNN.}
        \vspace{-5pt}
	\label{fig:design_architecture}
\end{figure}

To efficiently implement the proposed execution paradigm, we design a reconfigurable hardware accelerator, termed TVL-HGNN, with its overall architecture shown in Fig.~\ref{fig:design_architecture}. At the system level, it employs a multi-channel design that enables parallel processing across multiple vertex groups. The computational module adopts a unified architecture that supports dynamic resource reconfiguration, thereby enhancing computational resource utilization. Additionally, the storage subsystem features a two-level cache hierarchy comprising globally shared and channel-private caches, designed to maximize on-chip data reuse and further improve memory efficiency.

\subsubsection{Hardware Components}

The hardware architecture of TVL-HGNN primarily consists of three subsystems: the computing subsystem, memory subsystem and control subsystem.

The computing subsystem is composed of three parts: the \textit{Vertex Grouper}, the \textit{Computing Module}, and the \textit{Activation Module}. The \textit{Vertex Grouper} is utilized to group target vertices based on cross-semantic neighborhood overlap, ensuring high data locality within each group, which will be elaborated in Section~\ref{sec:vertex_grouping}. The \textit{Computing Module} handles the main execution of HGNN inference within each group. It includes a set of novel reconfigurable processing elements (RPEs), of which each is structured as a reduction tree and will be detailed in Section~\ref{sec:micro_RPEs}. RPEs can be dynamically configured into either linear transformation mode or aggregation mode based on the computational requirements to improve resource utilization. The \textit{Activation Module} is responsible for applying nonlinear functions such as \textit{LeakyReLU}.

The memory subsystem primarily comprises a two-level \textit{Feature Cache}, consisting of global and local caches, which are used to cache the projected vertex feature vectors and intermediate aggregation results. Note that these caches are essentially lightweight cache-like buffers, indexed by vertex type, vertex identifier (ID), and execution stage ID, and employ a first-in-first-out replacement policy. In addition, a set of \textit{Buffer} units is dedicated to storing various forms of temporary data, including target vertices and their adjacency information, pre-trained weight parameters, and reusable attention parameters. The \textit{High Bandwidth Memory} (HBM) is responsible for storing the original graph structure and raw vertex features.

The control subsystem includes a global logic \textit{Scheduler} responsible for orchestrating overall task execution. A local \textit{Dispatcher} manages the assignment of computation tasks within specific computing modules, while the \textit{Memory Controller} handles the scheduling of data transfers between off-chip and on-chip memory.

\subsubsection{Reconfigurable PEs}
\label{sec:micro_RPEs}

\begin{figure}[!htbp] 
	\centering
	\vspace{-7pt}
	\includegraphics[width=0.48\textwidth]{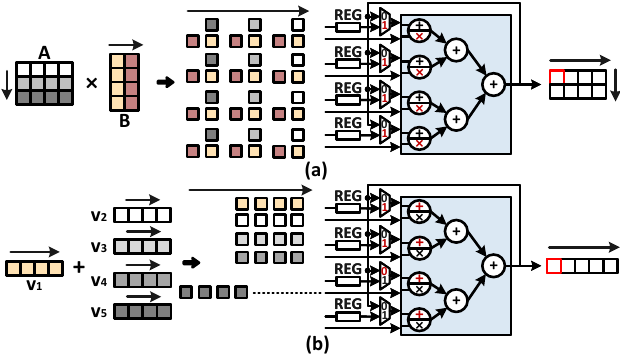}
	\caption{Micro-architecture and execution mode of RPEs: (a) Linear transformation mode; (b) Aggregation mode.}
        \vspace{-7pt}
	\label{fig:design_rpe}
\end{figure}

Fig.~\ref{fig:design_rpe} illustrates the microarchitecture of the RPEs designed for semantics-complete inference, along with its two applicable execution modes. Each RPE is a reconfigurable reduction tree, with the first layer consisting of multiply-or-accumulate (MOA) units and all subsequent layers composed of adders. As shown in Fig.~\ref{fig:design_rpe}(a), the \textbf{linear transformation mode} is primarily responsible for executing matrix multiplication operations, which are used in tasks such as feature projection and attention computation. For a matrix multiplication $A \times B$, the workload is mapped by decomposing the vectors element-wise, such that $A_{1,1}$ and $B_{1,1}$ are mapped to the first MOA unit, $A_{1,2}$ and $B_{2,1}$ to the second unit, and so on. Since each row of matrix $A$ must be multiplied and accumulated with all columns of matrix $B$, one end of MOA units receives input through a register to hold the operand from matrix $A$ constant, thereby reducing memory accesses for $A$. As shown in Fig.~\ref{fig:design_rpe}(b), the \textbf{aggregation mode} is mainly responsible for element-wise reduction over neighbor features. For an aggregation workload $v_1 \gets \{v_2, v_3, v_4, v_5\}$, the workload is mapped in a vector-wise manner: $v_1$ and $v_2$ are mapped to the first MOA unit, $v_3$ and $v_4$ to the second. For the unpaired odd vector $v_5$, the intermediate aggregation result is fed back to the input, and $v_5$ is delayed by three execution cycles so it can be aggregated with the prior aggregation result.

RPEs are designed to maximize the reuse of a unified set of reconfigurable hardware units for diverse operations, reducing design overhead and enhancing resource utilization. After the FP stage, most RPEs in a specific channel can be reconfigured into aggregation mode to support concurrent aggregation of numerous features, sustaining high throughput during semantics-complete inference. Moreover, the scheduling of operand buffering and result write-back also minimizes redundant memory accesses to operator matrices and intermediate results, further boosting overall efficiency.





\subsection{Overlap-driven Vertex Grouping}
\label{sec:vertex_grouping}

The semantics-complete execution paradigm treats each target vertex and its neighbors across all semantics as a basic aggregation workload block, creating an opportunity to group target vertices based on cross-semantic neighborhood overlap in order to enhance data locality. We first model an overlap-based hypergraph from the perspective of target vertices, then introduce an overlap-driven grouping method that maps vertices with high neighbor overlap to the same computing module. Finally, we present the microarchitecture design.





\subsubsection{Hypergraph Modeling}

\begin{figure}[!ht] 
	\centering
	\vspace{-8pt}
	\includegraphics[width=0.48\textwidth]{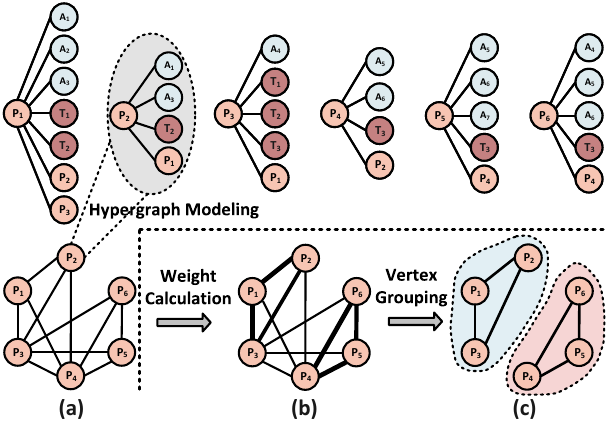}
	\caption{A running example of vertex grouping.}
        \vspace{-5pt}
	\label{fig:design_group_partition}
\end{figure}

As shown in Fig.~\ref{fig:design_group_partition}(a), we model the aggregation load of each target vertex, including the vertex itself and its neighbors across all semantics, as a super vertex. To capture neighborhood overlap under heterogeneous relations, we construct weighted edges between pairs of super vertices. Specifically, an edge is introduced if two super vertices share neighbors in the original HetG, with the edge weight $w_o$ defined by the \textit{Jaccard Similarity} of their neighborhoods: $w_o=\frac{|N(v_i) \cap N(v_j)|}{|N(v_i) \cup N(v_j)|}$, where $N(v_i)$ and $N(v_j)$ denote the multi-semantic neighborhoods of target vertices $v_i$ and $v_j$, respectively, including themselves. This modeling results in a hypergraph where each super vertex encapsulates the complete aggregation workload of a target vertex, and edge weights quantify the potential for shared neighbor reuse as illustrated in Fig.~\ref{fig:design_group_partition}(b), in which thicker edges indicate stronger overlap. This modeling is applied only to the top 15\% of high-degree target vertices, which already cover most neighboring vertices due to the power-law distribution, while low-degree vertices are grouped using a simple sequential strategy.

\subsubsection{Vertex Grouping}


Building on the modeling above, the problem of assigning target vertices to processing channels to maximize neighbor reuse is cast as a community detection task. Specifically, the objective is to group vertices linked by high-weight edges into the same community, as shown in Fig.~\ref{fig:design_group_partition}(c), with inter-community edge weights relatively lower. Inspired by the \textit{Louvain} algorithm~\cite{louvain}, we propose an overlap-driven vertex grouping method tailored to the multi-channel hardware architecture, as detailed in Algorithm~\ref{alg:design_vertex_group}. This method extends the standard \textit{Louvain} approach by integrating the proposed neighborhood overlap metric $w_o$ into the modularity calculation. Additionally, we enhance its applicability by using a streaming group generation workflow, enabling pipelined execution between group generation and processing.

\begin{algorithm}[!t]
\SetAlgoLined
\caption{Overlap-driven Vertex Grouping}
\label{alg:design_vertex_group}
\SetKwInOut{Input}{Input}
\SetKwInOut{Output}{Output}
\SetKwInOut{Initial}{Initial}

\Input{Hypergraph $G_H = (V_H, E_H)$, max size $N_{max}$}
\Output{Set of groups $C$}
\Initial{
    $C \gets \emptyset$, Unassigned vertices $U \gets V_H$ \\
}


\While{$U \neq \emptyset$}{
    Choose a seed vertex $v_s \in U$ \\
    Initialize new group $C_{new} \gets \{v_s\}$ \\
    Remove $v_s$ from $U$ \\

    \While{$|C_{new}| < N_{max}$}{
        $\Delta Q_{max} \gets 0$, $v^* \gets \phi$ \\
        
        \For{each $v \in \text{Neighbors}(C_{new}) \cap U$}{
            $\Delta Q \gets \text{ComputeModularityGain}(v, C_{new})$ \\
            \If{$\Delta Q > \Delta Q_{max}$}{
                $\Delta Q_{max} \gets \Delta Q$, $v^* \gets v$ \\
            }
        }

        \If{$\Delta Q_{max} > 0$}{
            Add $v^*$ to $C_{new}$ \\
            Remove $v^*$ from $U$ \\
        }
        \textbf{else} break
    }

    Add $C_{new}$ to $C$ \Comment{\textit{Can be sent for processing}} \\ 
}

\Return $C$
\end{algorithm}

Specifically, for the constructed hypergraph $G_H$, a random unvisited vertex is first selected as the seed vertex $v_s$, and a new group $C_{new}$ is initialized with $v_s$ (lines 2-3). For each target vertex $v$ connected to $C_{new}$, the modularity gain $\Delta Q$ is computed under the attempt of adding $v$ to $C_{new}$ as in \textit{Louvain} algorithm. The neighbor vertex $v^*$ that yields the maximum modularity gain is identified (lines 7-12). If the gain is positive, $v^*$ is added to $C_{new}$. Otherwise, it indicates that no further modularity improvement can be achieved by including additional vertices, and the generation of the next group begins (lines 13-19). To prevent load imbalance caused by oversized groups, an upper bound on the number of vertices per group is imposed (line 5), defined as the total number of target vertices divided by the number of parallel processing channels.

\begin{figure}[!ht] 
	\centering
	\vspace{-5pt}
	\includegraphics[width=0.48\textwidth]{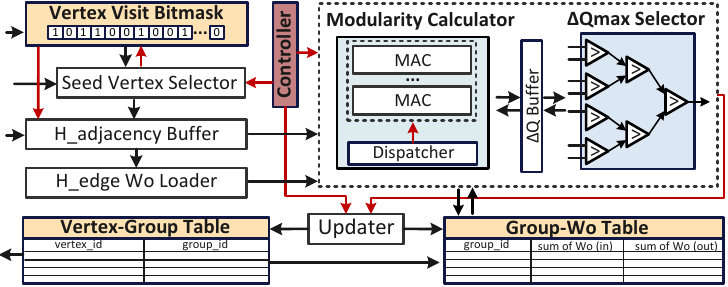}
	\caption{Microarchitecture of vertex grouper.}
        \vspace{-5pt}
	\label{fig:design_micro_grouper}
\end{figure}


The microarchitecture of the vertex grouper is illustrated in Fig.~\ref{fig:design_micro_grouper}. A \textit{Vertex Visit Bitmask} tracks which vertices have been visited. At the beginning of each group generation, the \textit{Seed Vertex Selector} selects an unvisited vertex as the seed $v_s$ to initialize a new group $C_{new}$. The \textit{H\_adjacency Buffer} stores unvisited neighbors of $v_s$, while the \textit{H\_edge Wo Loader} fetches pre-computed neighborhood overlap weights. For each neighbor $v$, the \textit{Modularity Calculator} computes the modularity gain $\Delta Q$ for adding $v$ to $C_{new}$ using multiply-and-accumulate (MAC) units. The \textit{$\Delta Q_{max}$ Selector} then identifies the neighbor with the highest gain utilizing a comparison tree. If $\Delta Q_{max}$ is positive, the \textit{Updater} updates both the \textit{Vertex-Group Table}, recording each vertex’s group ID, and the \textit{Group-Wo Table}, tracking intra- and inter-group weights. Otherwise, the grouper proceeds to generate the next group.

The proposed overlap-driven vertex grouping technique serves two key purposes. \textbf{First}, it partitions the workload into groups, enabling inter-group parallelism to enhance inference performance. \textbf{Second}, by grouping vertices with high neighbor overlap, it improves data locality and significantly reduces DRAM accesses through the reuse of neighboring features.

\section{Evaluation}
\label{sec:evaluation}

\subsection{Experiment Setup}

\textit{\textbf{Methodology.}} The performance and energy efficiency of TVL-HGNN are assessed utilizing the following tools.

\textit{Cycle-accurate Simulator.} 
We implement TLV-HGNN within a cycle-accurate simulator to assess its performance in terms of execution cycles, and integrate Ramulator~\cite{ramulator} to precisely model off-chip memory accesses to HBM.

\textit{CAD Tools.}
The Synopsys Design Compiler with the TSMC 12$nm$ standard VT library is employed for synthesizing the RTL implementation of each module. Power consumption is estimated using Synopsys PrimeTime PX. The module with the longest critical path delay measures 0.81 ns, enabling TLV-HGNN to reliably operate at a 1.0 GHz clock frequency.

\textit{Memory Measurements.}
The access latency, energy consumption, and area of on-chip memory components are estimated using Cacti 6.5~\cite{CACTI}. Four distinct scaling factors are applied to adjust these estimates to the 12nm technology node, following the approach in work~\cite{technology_scale}. The latency and energy consumption of HBM1.0 are simulated via Ramulator and estimated at 7 pJ/bit as in work~\cite{7pj}.

\textit{\textbf{Platforms.}}
Performance is evaluated on an NVIDIA A100 GPU with metrics collected via Nsight Compute, using Float32 precision. The SOTA HGNN accelerator HiHGNN~\cite{HiHGNN} is included as a baseline. Table~\ref{tb:evaluation_platform} summarizes the experimental setup. For fairness, the HBM capacity of HiHGNN and TVL-HGNN is aligned with that of the A100.

\begin{table}[!ht]
\vspace{-3pt}
\centering
\caption{Specifications of the platforms.} 
\vspace{-2pt}
\label{tb:evaluation_platform}
\renewcommand\arraystretch{1.2}
\setlength\tabcolsep{2pt}%
\resizebox{0.48\textwidth}{!}{
\begin{tabular}{ccccll}
\toprule
 & \textbf{A100} & \textbf{HiHGNN}   & \multicolumn{3}{c}{\textbf{TVL-HGNN}}                                                                                                 \\ \midrule
\textbf{\begin{tabular}[c]{@{}c@{}}Peak\\ Performance\end{tabular}}  & \begin{tabular}[c]{@{}c@{}} 19.5 TFLOPS, \\  1.41 GHz \end{tabular}                 & \begin{tabular}[c]{@{}c@{}} 16.38 TFLOPS, \\  1.0 GHz \end{tabular}                & \multicolumn{3}{c}{\begin{tabular}[c]{@{}c@{}} 15.36 TFLOPS, \\  1.0 GHz \end{tabular}} \\ \midrule
\textbf{\begin{tabular}[c]{@{}c@{}}On-chip\\ Memory\end{tabular}}  & \begin{tabular}[c]{@{}c@{}}20 MB (L1 Cache),\\40 MB (L2 Cache) \end{tabular} & \begin{tabular}[c]{@{}c@{}}2.44 MB (FP-Buf),\\14.52 MB (NA-Buf),\\0.12 MB (SA-Buf),\\ 0.38 MB (Att-Buf)\end{tabular} & \multicolumn{3}{c}{\begin{tabular}[c]{@{}c@{}}1.64 MB (Weight Buffer),\\0.60 MB (Target Buffer),\\1.00 MB (Attention Buffer), \\ 1.40 MB (Adjacency Buffer) , \\ 1.20 MB (Grouper Buffers), \\ 6.00 MB (Feature Cache)\end{tabular}} \\ \midrule
\textbf{\begin{tabular}[c]{@{}c@{}}Off-chip\\ Memory\end{tabular}} & \begin{tabular}[c]{@{}c@{}}2039 GB/s, \\80 GB, HBM2e\end{tabular}                                & \begin{tabular}[c]{@{}c@{}}512 GB/s, \\80 GB, HBM1.0\end{tabular}                                & \multicolumn{3}{c}{\begin{tabular}[c]{@{}c@{}}512 GB/s, \\80 GB, HBM1.0\end{tabular}}                                                                                                                            \\ \bottomrule
\end{tabular}
}
\scriptsize
\vspace{-3pt}
\end{table}

\textit{\textbf{Benchmarks.}}
We evaluate three representative HGNN models: RGCN~\cite{R-GCN}, RGAT~\cite{R-GAT}, and NARS~\cite{NARS}, all implemented using DGL 1.0.2~\cite{DGL} and trained with the hyperparameters specified in their original papers. Experiments are conducted on five widely used datasets from~\cite{openhgnn}: ACM, IMDB, DBLP, AM, and Freebase (FB). The first three are small-scale and adopted in HiHGNN~\cite{HiHGNN} for fair comparison, while the latter two are significantly larger, with up to two orders of magnitude more vertices, edges, and semantics.



\subsection{Overall Results}

\begin{figure*}[!htbp] 
    \vspace{-10pt}
    \small
    \centering
    \includegraphics[width=1.0\textwidth]{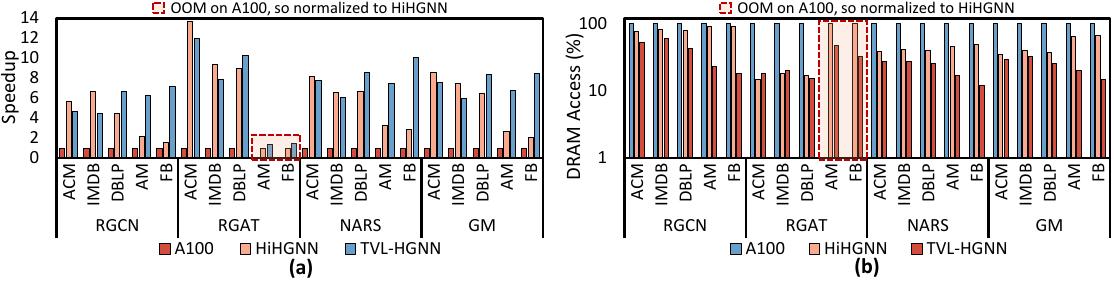}
    \vspace{-14pt}
    \caption{Overall results: (a) Speedup; (b) DRAM Access.}
    \label{fig:evaluation_speedup_dram}
    \vspace{-15pt}
\end{figure*}




\subsubsection{Speedup} 

As shown in Fig.~\ref{fig:evaluation_speedup_dram}(a), TVL-HGNN achieves average speedups of 7.85$\times$ over the A100 GPU and 1.41$\times$ over HiHGNN. For cases where A100 encounters OOM, performance is normalized to HiHGNN. The gain over HiHGNN mainly stems from reconfigurable RPEs that improve resource utilization and overlap-driven vertex grouping that enhances data locality and reduces DRAM accesses during the NA stage. Compared to A100, additional benefits arise from the customized architecture and optimized data pathways.



\subsubsection{Memory Efficiency}

Table~\ref{tb:evaluation_memory_expansion} reports memory expansion ratios on the AM dataset, with similar trends across others. TVL-HGNN significantly mitigates memory expansion via the semantics-complete paradigm, avoiding OOM issues and maintaining low expansion ratios, thereby improving scalability on large graphs without significantly compromising inference performance as batch-wise execution does.

\begin{table}[!ht]
\vspace{-4pt}
 \caption{Memory expansion ratios on AM dataset.} 
 \vspace{-2pt}
 \label{tb:evaluation_memory_expansion}
 \centering
 \renewcommand\arraystretch{1.0}
    \resizebox{0.36\textwidth}{!}{
\begin{tabular}{|c|c|c|c|}
\hline
\textbf{Model} & \textbf{A100} & \textbf{HiHGNN} & \textbf{TVL-HGNN} \\ \hline
RGCN   & 14.76                 & 8.21  & 1.64     \\ \hline
RGAT   & OOM                 & 18.27  & 2.38     \\ \hline
NARS & 13.64                 & 7.52  & 1.59     \\ \hline
\end{tabular}
}
\vspace{-4pt}
\end{table}

As presented in Fig.~\ref{fig:evaluation_speedup_dram}(b), TVL-HGNN reduces DRAM access by 76.46\% compared to A100 and 49.63\% compared to HiHGNN on average. This reduction mainly results from the semantics-complete execution paradigm that eliminates the need to store or reload intermediate aggregations while avoiding redundant access to target features, and the overlap-driven vertex grouping technique that enhances data locality and minimizes repeated neighboring feature accesses.


\subsubsection{Energy and Its Breakdown} 

\begin{figure}[!hptb] 
    \vspace{-8pt}
    \small
    \centering
    \includegraphics[width=0.48\textwidth]{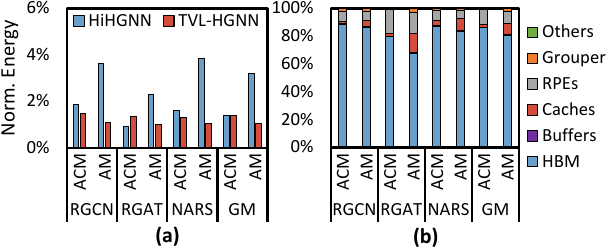}
    \vspace{-6pt}
    \caption{Energy and its breakdown on AM dataset.}
    \label{fig:evaluation_energy}
    \vspace{-5pt}
\end{figure}

For clarity, we present energy consumption results on a representative small dataset (ACM) and a large dataset (AM). As shown in Fig.~\ref{fig:evaluation_energy}(a), TVL-HGNN reduces energy consumption by 98.79\% over A100 and 32.61\% over HiHGNN on average, primarily driven by a significant reduction in DRAM accesses. Fig.~\ref{fig:evaluation_energy}(b) shows the energy breakdown of TVL-HGNN, where off-chip DRAM access constitutes the majority of energy usage, followed by the RPEs responsible for the primary computational workload.

\subsubsection{Variation Across Models and Datasets}

On \textbf{model level}, RGAT’s multi-head attention mechanism in NA stage introduces additional redundant neighbor accesses, enabling TVL-HGNN to achieve the highest speedups along with the greatest reductions in DRAM accesses and energy consumption over A100. However, HiHGNN's bitmap-based attention reuse mitigates redundancy, making this trend reversed. On \textbf{dataset level}, large graphs with higher edge-to-vertex ratios introduce more redundancy, while their longer inference times amortize the grouping overhead. Thus, compared to both A100 and HiHGNN, TVL-HGNN yields greater performance, memory, and energy benefits on larger datasets than smaller ones. \textbf{Notably}, while TVL-HGNN underperforms HiHGNN slightly on small datasets, it consistently delivers superior performance on large graphs, achieving up to 4.62$\times$ speedup.

\subsubsection{Area and Power}
As shown in Table~\ref{tb:evaluation_chip_area_power}, TVL-HGNN integrates 11.84 MB on-chip SRAM, 2048 RPEs across four channels, 512 MAC units for vertex grouper, and control logic, resulting in a total chip area of 16.56~$mm^2$ and power consumption of 10.61~W. On-chip memory occupies 47.33\% of the area and 8.34\% of the power, while the computing module dominate with 43.11\% area and 82.73\% power. Compared to HiHGNN, TVL-HGNN delivers higher scalability and performance with lower area and power overhead.

\begin{table}[!ht]
 \caption{Characteristics of TVL-HGNN (TSMC 12 $nm$).} 
 \label{tb:evaluation_chip_area_power}
 \centering
 \renewcommand\arraystretch{1.0}
    \resizebox{0.49\textwidth}{!}{
\begin{tabular}{|l|l|r|r|r|r|}
\hline
\multicolumn{2}{|l|}{\begin{tabular}[c]{@{}c@{}}\textbf{Component or Block}\end{tabular}} & \begin{tabular}[r]{@{}c@{}}\textbf{Area ($mm^2$)}\end{tabular} & \textbf{\%} & \begin{tabular}[r]{@{}c@{}}\textbf{Power ($mW$)}\end{tabular}  & \textbf{\%} \\ \hline \hline
\multicolumn{2}{|l|}{{TVL-HGNN (4 Channels)}}    &16.56 &100 &10613.71 &100    \\ \hline \hline
\multicolumn{6}{|c|}{\begin{tabular}[c]{@{}c@{}} \textbf{Breakdown by Functional Block} \end{tabular}} \\ \hline
\multicolumn{2}{|l|}{Feature Caches}   &4.42 &26.69 &498.93 &4.70    \\
\multicolumn{2}{|l|}{On-chip Buffers}   &3.42 &20.64 &385.84 &3.64     \\
\multicolumn{2}{|l|}{Computing Module}   &7.14 &43.11 &8780.80 &82.73    \\
\multicolumn{2}{|l|}{Activation Module}   &0.11 &0.64 &156.80 &1.48    \\
\multicolumn{2}{|l|}{Vertex Grouper}    &1.39 &8.42 &726.99 &6.84  \\
\multicolumn{2}{|l|}{Others}          &0.08 &0.50 & 64.35 & 0.61    \\  \hline
\end{tabular}
}
\end{table}

\subsection{Effects of Optimizations}

In this section, we adopt an incremental evaluation across multiple models under AM dataset to demonstrate the effectiveness of the proposed optimizations. The \textbf{-B} version is a single-channel TVL-HGNN using the conventional per-semantic execution without vertex grouping. The \textbf{-S} version incorporates our semantics-complete execution paradigm. The \textbf{-P} version further extends TVL-HGNN to four channels with random vertex grouping. Finally, the \textbf{-O} version applies our neighborhood overlap-driven vertex grouping method.

\subsubsection{Effects of Semantics-complete Execution Paradigm}

\begin{figure}[!hptb] 
    \vspace{-10pt}
    \small
    \centering
    \includegraphics[width=0.48\textwidth]{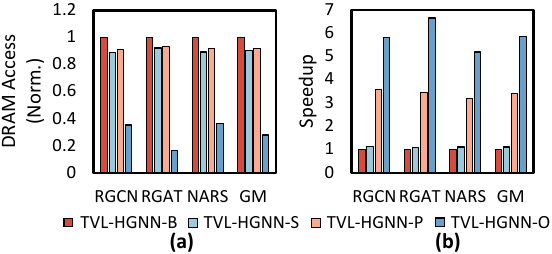}
    \vspace{-3pt}
    \caption{Effects of optimizations on AM dataset: (a) Number of DRAM Accesses; (b) Speedup.}
    \label{fig:optimization_effects}
    \vspace{-8pt}
\end{figure}

The semantics-complete execution paradigm eliminates storage and access of intermediate aggregation results, as well as redundant accesses to target vertex features. Compared to the \textbf{-B} version, \textbf{-S} reduces DRAM access by 9.82\% on average across models as presented in Fig.~\ref{fig:optimization_effects}(a), yielding a 1.11$\times$ speedup. It also alleviates memory expansion significantly, as reported in Table~\ref{tb:evaluation_memory_expansion}, enabling efficient support for larger-scale datasets.

\subsubsection{Effects of Overlap-driven Vertex Grouping}

As illustrated in Fig.~\ref{fig:optimization_effects}(a), the overlap-driven vertex grouping method employed in the \textbf{-O} configuration helps to reduce DRAM accesses by an average of 66.95\% compared to \textbf{-P}, yielding a 1.72$\times$ performance boost as shown in Fig.~\ref{fig:optimization_effects}(b). Notably, the \textbf{-O} configuration achieves a substantial 5.29$\times$ performance improvement over \textbf{-S}. This significant gain is attributed not only to the reduction in DRAM access but also to the effective exploitation of inter-group parallelism as the same in \textbf{-P}.



\section{Related Work}

\textit{GNN Accelerators:} GNN accelerators have garnered significant interest from the architecture community in recent years~\cite{HyGCN, igcn, ReGNN, FlowGNN, GCNTrain, PANG}. A small subset of these works~\cite{igcn, ReGNN} addresses redundant memory accesses in GNNs; however, their methods are fundamentally ill-suited for HGNN optimization due to the multi-semantic nature of HetGs. For example, \textit{I-GCN}~\cite{igcn} adopts a graph-traversal-based strategy to partition HomoGs into densely connected islands, supported by dedicated hardware. This approach, however, encounters two major limitations in the HGNN context. \textbf{First}, semantic graphs in HGNNs are typically bipartite, lacking direct connections among target vertices, which undermines the effectiveness of connectivity-based partitioning. \textbf{Second}, the presence of multiple semantic graphs with diverse locality patterns introduces substantial overhead when each graph is partitioned independently, especially at large scales. In addition, existing GNN accelerators are also ineffective in addressing the memory expansion problem unique to HGNNs.


\textit{HGNN Accelerators:} Only a limited number of work~\cite{MetaNMP, HiHGNN, ADE-HGNN} have focused on inference acceleration for emerging HGNNs. \textit{HiHGNN}~\cite{HiHGNN} proposes a bound-aware stage fusion approach to enable parallel execution of different inference stages, along with a scheduling method based on semantic graph similarity to maximize data reuse across semantic graphs. \textit{ADE-HGNN}~\cite{ADE-HGNN} introduces a neighbor pruning strategy guided by attention importance to reduce the computational load of the NA stage, and employs fine-grained pipelining to amortize pruning overhead. \textbf{However}, existing works have largely overlooked the issues of memory expansion and redundant memory accesses inherent in the current execution paradigm during HGNN inference, resulting in limited scalability and suboptimal efficiency.

\section{Conclusion}

This work first proposes a semantics-complete inference paradigm and develops a multi-channel reconfigurable accelerator for efficient execution. Additionally, it introduces a neighborhood-overlap-based vertex grouping method that exploits inter-group parallelism and maximizes intra-group neighbor reuse, reducing off-chip memory access. Extensive experimental results demonstrate the superior performance of TVL-HGNN over GPU A100 and SOTA HGNN accelerator.

\bibliographystyle{IEEEtran}
\bibliography{refs}

\end{document}